\documentclass[conference]{IEEEtran}
\IEEEoverridecommandlockouts
\usepackage{cite}
\usepackage{amsmath,amssymb,amsfonts}
\usepackage{algorithmic}
\usepackage{graphicx}
\usepackage{textcomp}
\usepackage{xcolor}

\usepackage{multirow}
\usepackage{makecell}
\usepackage{pbox}
\usepackage{breakurl}
\usepackage{url}

\def\BibTeX{{\rm B\kern-.05em{\sc i\kern-.025em b}\kern-.08em
    T\kern-.1667em\lower.7ex\hbox{E}\kern-.125emX}}
        
\usepackage{fancyhdr}
\newcommand{\emojiwidth}{0.025\textwidth}

\begin{document}

\title{Towards Understanding Emotional Intelligence for Behavior Change Chatbots}

\author{\IEEEauthorblockN{Asma Ghandeharioun}
\IEEEauthorblockA{\textit{MIT Media Lab}\\
Cambridge, MA, US \\
asma\_gh@mit.edu}
\and
\IEEEauthorblockN{Daniel McDuff}
\IEEEauthorblockA{\textit{Microsoft Research}\\
Redmond, WA, US \\
damcduff@microsoft.com}
\and
\IEEEauthorblockN{Mary Czerwinski}
\IEEEauthorblockA{\textit{Microsoft Research}\\
Redmond, WA, US \\
marycz@microsoft.com}
\and
\IEEEauthorblockN{Kael Rowan}
\IEEEauthorblockA{\textit{Microsoft Research}\\
Redmond, WA, US \\
kael.rowan@microsoft.com}
}

\maketitle

\begin{abstract}
A natural conversational interface that allows longitudinal symptom tracking would be extremely valuable in health/wellness applications. However, the task of designing emotionally-aware agents for behavior change is still poorly understood. In this paper, we present the design and evaluation of an emotion-aware chatbot that conducts experience sampling in an empathetic manner. We evaluate it through a human-subject experiment with N=39 participants over the course of a week. Our results show that extraverts preferred the emotion-aware chatbot significantly more than introverts. Also, participants reported a higher percentage of positive mood reports when interacting with the empathetic bot. Finally, we provide guidelines for the design of emotion-aware chatbots for potential use in mHealth contexts.
\end{abstract}


\section{Introduction}

The Experience Sampling Method (ESM) is a technique in which feelings or activities are recorded at the moment, either at randomly selected or predefined times \cite{prescott1981environmental}. ESM is less influenced by memory-bias, compared to retrospective self-reports. It has shown promise in understanding affect in context, linking it to events, and unraveling its temporal patterns \cite{scollon2009experience}. ESM has proved to be valid and reliable \cite{csikszentmihalyi2014validity}, and particularly useful in symptom tracking in psychosomatic medicine \cite{conner2012trends}. There have been multiple efforts for streamlining ESM (e.g. \cite{ghandeharioun2016kind}) and improving user engagement (e.g. \cite{taylor2019use}). However, other aspects of ESM design, such as delivery via an agent, have not been fully studied.

ESM provides a means for self-reflection and both unmediated and technology-mediated self-reflection have been shown to improve wellbeing \cite{isaacs2013echoes}. 
In this paper, we aim to explore if we can further improve the positive effects of self-reflection on behavior change and one's sense of wellbeing by delivering ESM through an emotionally sentient agent. More specifically, we study the influence of interaction with an emotion-aware chatbot on one's mood.

Nowadays, virtual agents (VA) have been successfully deployed in multiple settings, ranging from education \cite{d2007toward} to healthcare \cite{devault2014simsensei, ring2016affectively}, particularly in symptom tracking. While there is a strong focus on building applications to assess health, there is scientific evidence that making such applications empathetic plays a significant role in their acceptance and success and improves user experience \cite{liu2005embedded}. An agent that is adaptive to the user's emotional state is perceived as more trustworthy, valuable, and intelligent \cite{bickmore2001relational,gratch2007creating,lucas2014s}. 

However, when the context is more personal and nuanced, such as when the agent asks about the user's mood or mental wellbeing, there are further intricate details that need to be studied. What does the concept of emotional intelligence for an automated personal assistant mean in such contexts? Should the technology be designed for replicating human experience and emotional expressiveness? Or do users feel more comfortable opening up to a neutral and objective assistant? Does personality type influence one's preference for affective personal assistants? Does interaction with an emotionally-aware agent influence users' behavior in and of itself?

To address these questions, we designed and implemented an emotion-aware chatbot for the general population. This chatbot conducts experience sampling in an empathetic manner; this means that it is not only the instrument to capture self-reports, but also responds with emotionally appropriate conversations. It acknowledges the user's emotional state, similar to what an empathetic companion would do. We evaluated different aspects of this emotion-aware chatbot through a week-long randomized controlled trial with N=39 participants where we compared an emotionally appropriate condition (Emotion-Aware) to a neutral condition (Control). Our results showed that participants recorded a higher percentage of positive emotions using the empathetic bot compared to the neutral bot. Also, extraverts preferred the emotionally expressive bot significantly more than introverts. We present future directions of how such a chatbot can be used for longitudinal symptom tracking and for delivering mHealth interventions.

\section{Related Work}\label{sec:relatedWork}

The experience sampling method (ESM) \cite{prescott1981environmental} was born in 1970s with the advent of pagers. It is a validated technique used for capturing frequency, intensity, and overall patterns of one's emotional experience \cite{csikszentmihalyi2014validity}. ESM alleviates people's inability to provide accurate retrospective information on their daily behavior and experience by capturing such information in the moment \cite{bernard1984problem}.

Despite the strengths of ESM, it also has limitations. This method puts heavy demand on research participants, which makes it a better fit for conscientious individuals rather than the general population \cite{scollon2009experience}. There have been multiple efforts in improving the average user's engagement and compliance. Some researchers have tried to reduce the burden on the user by incorporating ESM more seamlessly into their daily pipelines, for example, by placing it in the phone unlock screen \cite{ghandeharioun2016kind}. Other researchers have designed engaging games, making short questionnaires part of the game flow, and have validated ESM responses captured in the game in comparison to the traditional setting \cite{taylor2019use}.

While these efforts have tried to address the issue of user engagement with ESM, the tone of delivery of ESM and its implications are not well studied. Specifically, ESM goes beyond being purely a method for capturing data. It provides a means for reflecting on one's past more objectively, which has the potential to improve psychological wellbeing and personal growth \cite{bryant2005using}. ESM is widely used in applications for improving mental health and wellbeing. Recently, personal assistants, chatbots, and virtual assistants (VAs) are are being used more often for conducting experience sampling. Therefore, it becomes imperative to study the characteristics of such an agent delivering ESM.

 In health and mental health contexts, patients are sometimes reluctant to respond honestly. Extensive research on VAs shows that people are more comfortable disclosing health symptoms to VAs when they know they are automated and there is no human behind the scenes \cite{lucas2014s}. Interestingly, when VAs exhibit human qualities, they could further improve the relationship with the user. VAs are better at creating rapport when contingent on the human's responses \cite{gratch2007creating}. Also, they are more successful in establishing trust when using relational conversational strategies \cite{bickmore2001relational}. It remains an open question how emotional-awareness and empathy in an agent which conducts mood experience sampling is perceived by users, if it is mediated by the user's personality, and if it affects the user's mood.

\section{System Design}\label{sec:systemDesign}

\begin{figure}[t!]
  	\includegraphics[width=1.0\columnwidth]{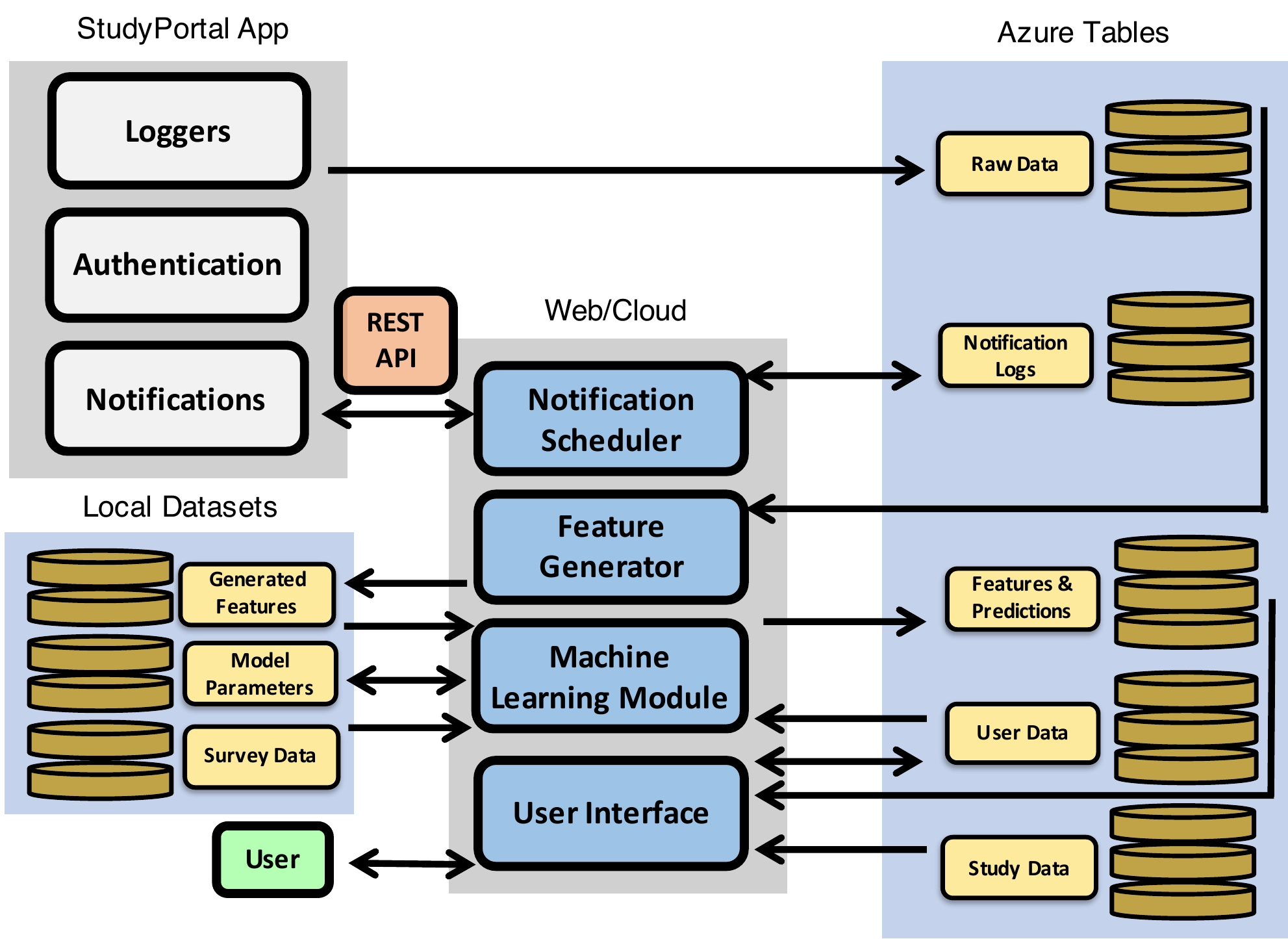}
  \caption{System design}
   \label{fig:system}
\end{figure}

We designed a conversational bot interface to conduct experience sampling. It specifically samples user's mood and is crafted to respond to it appropriately. For example, it responds positively to an expression of excitement, while responding sympathetically to an expression of stress. In this section we describe the system design in detail.

\subsection{Mobile Application}

The mobile application administers experience sampling and uses affective accompanying text. The app adjusts its behavior based on the group condition. Fig.~\ref{fig:system} depicts the system design.

\begin{figure}[t!]
	\centering
   	\includegraphics[width=0.75\columnwidth]{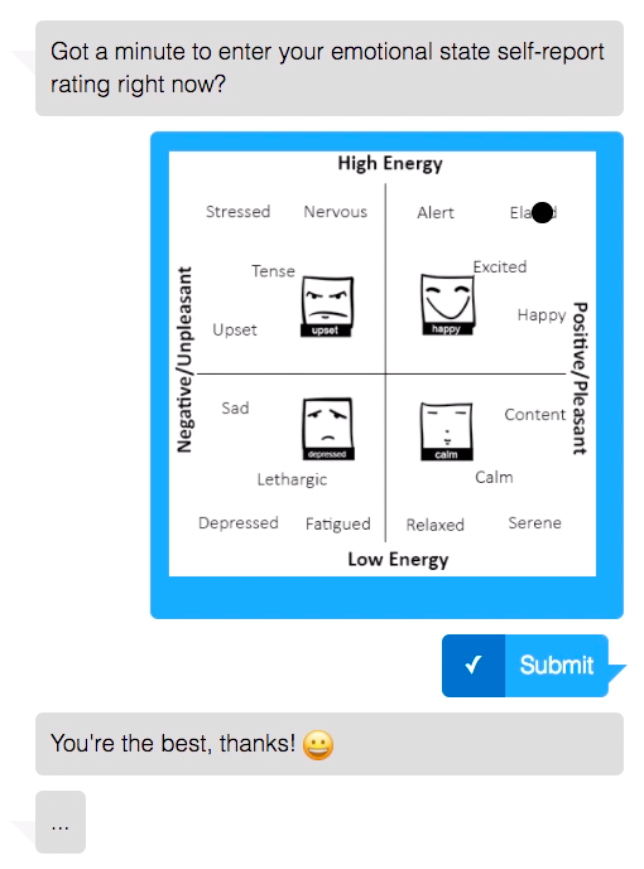}
   	\caption{The visual design of the user interface. The bubbles are color coded to show if they are coming from the agent (gray) or have already been answered (blue).}
    	\label{fig:UI}
        \vspace{-0.5cm}
 \end{figure}

The mobile app consists of a web-based user interface (UI) (Fig.~\ref{fig:UI}). The UI visualizes the conversations between the agent and the user. The content appears within bubbles that are left- or right-aligned based on the speaker (human or agent). Also, the bubbles are color coded to show if they are coming from the agent (gray), are prompts for the user to respond to (green), or have already been answered and are no longer editable (blue). To make the experience more realistic, the agent starts typing for one second before the text appears (see Fig.~\ref{fig:UI}). The content is selected from the pool of scripted texts by a rule-based decision tree according to the group condition and the user's most recently recognized affective state.

The web-based UI is built upon the StudyPortal platform which is designed to handle different OS types \cite{rowan2013studyportal}. In our case, StudyPortal is in charge of delivering notifications to the participants' phones and considering their history of previous responses.

\subsection{Experience Sampling}
To make the chatbot design emotionally intelligent, it needs to reason about the user's current affective state \cite{jeong2016improving}. To capture ground-truth emotion labels, we administered experience sampling five times a day and explicitly asked the participants to rate their mood. We adopted Russel's two-dimensional model of emotion \cite{russell1980circumplex} as our primary ``gold-standard'' mood measurement technique. This is one of the most prevalent and highly cited models of emotion and considers two dominant dimensions for mood: valence (pleasure - displeasure) and arousal (high energy - low energy). Horizontal and vertical axes correspond to valence and arousal respectively. To make it easier for users to self-report their mood, we included sample icons (visual cues) and emotional states (textual cues) that fall under the corresponding quadrants (See the experience sampling grid in Fig.~\ref{fig:UI}). This visual grid captures continuous values between 0 and 1 for both valence and arousal.

\subsection{Agent Dialog/Communications}
\label{sec:scripted-text}

\begin{figure}[h!]
	\centering
   	\includegraphics[width = 1.0\columnwidth]{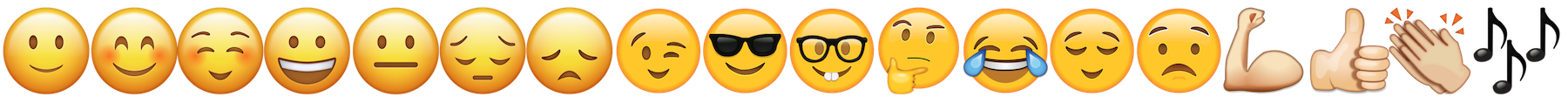}
   	\caption{Set of emojis used.}
    	\label{fig:emojis}
 \end{figure}

For smooth communications between the agent and the user, we scripted dialog that was emotionally expressive and added emojis (from the set depicted in Fig.~\ref{fig:emojis}) when appropriate to better communicate emotions. In the emotional condition, each textual interaction had an average of 1.3 emojis, where there was an emoji per 6.5 words. In order to keep the content more realistic and engaging, we have scripted 6 different phrasings for each dialog interaction and randomly selected one when starting a conversation. For the Control condition, we scripted similar texts, but so as to be completely neutral without any expression of affect or use of emojis. Table \ref{tab:text} provides an example of the affective vs. neutral text when the user has just responded to an ESM prompt.

  \begin{table}[t!]
  \caption{An example of affective vs. neutral textual content after receiving a mood report. TL, TR, BL and BR refer to the spatial locations on the 2x2 Russell circumplex model of emotion, e.g. TL refers to Top Left quadrant.}
   \centering
   \begin{tabular}{p{1.5cm}  p{6cm} }
     \textbf{Condition} & \textbf{Content} \\
      Neutral & Thanks for the rating.\\
     TL & Oh I see. I'm sorry to hear that. \includegraphics[width = \emojiwidth]{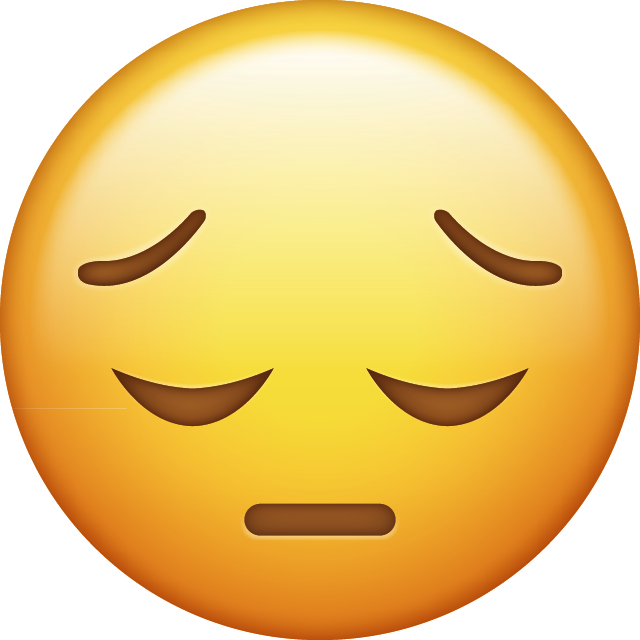}\\
     TR & I'm so proud of you \includegraphics[width = \emojiwidth]{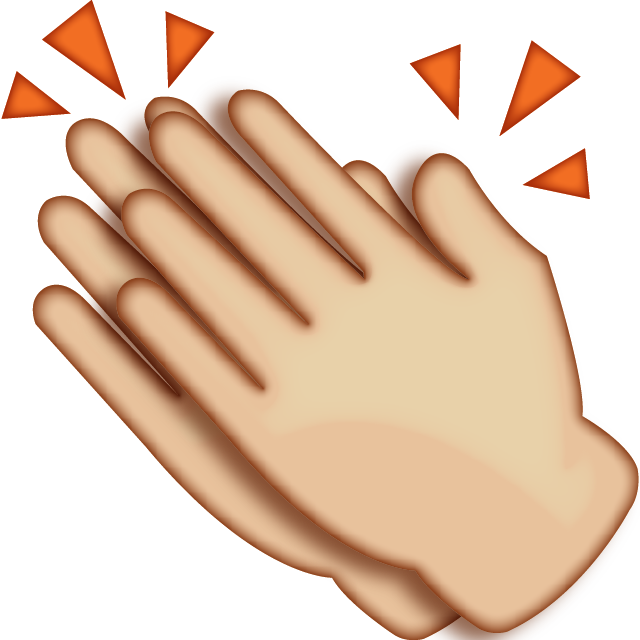} , thanks! \includegraphics[width = \emojiwidth]{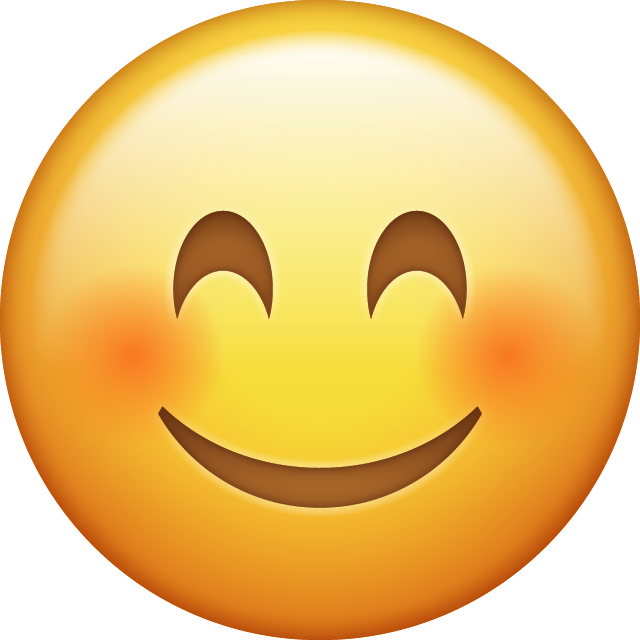} \\
     BL & So sorry \includegraphics[width = \emojiwidth]{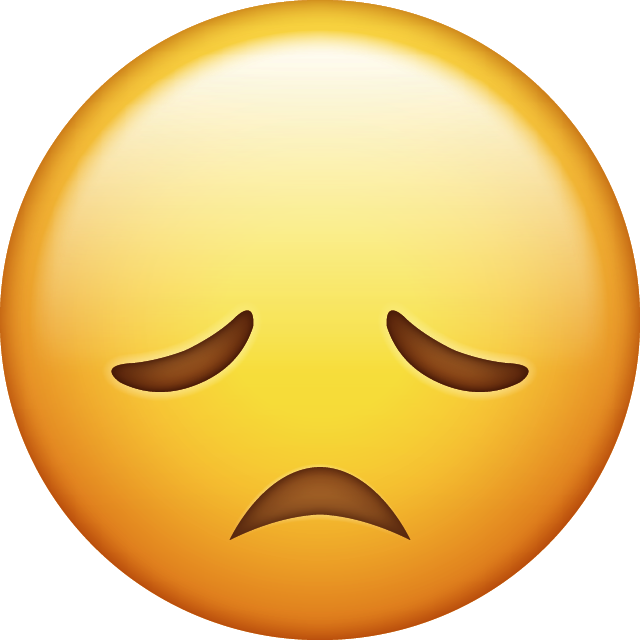}, Hang in there and thanks. \\
     BR & Nice. Thanks! \includegraphics[width = \emojiwidth]{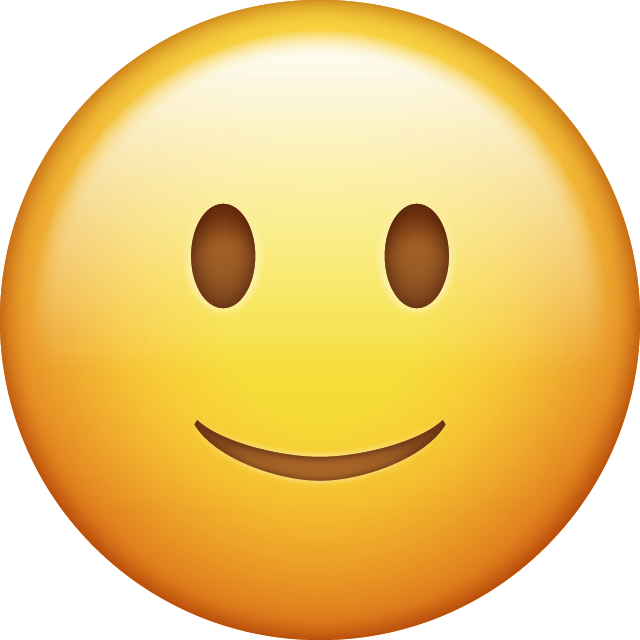} \\
   \end{tabular}
     \label{tab:text}
\end{table}

\section{Human Subjects}\label{sec:humanSubjects}

The study protocol was approved by the institutional review board at [anonymous institution]. Participants signed-up for the study online and were randomly assigned to the Emotion-Aware condition or the Control group. Forty one participants were recruited. One participant dropped out early due the app's phone battery usage. Another participant had previous knowledge about the hypotheses of the study and thus was excluded. Among N=39 participants that completed the protocol successfully, 7 were females and 32 were males. This included 17 full-time employees (FTE), 17 interns, and 5 external members or contractors. 
The participants' ages ranged between 16 and 49 (M=29.4, SD=7.9). Gift-card raffles were held at this week-long experiment for 50 dollars. Among active users, three were randomly selected as winners of the raffle \footnote{This study was part of a larger experiment that lasted several weeks. Participants received \$200 gift-cards for successfully completing all the experiments.}. 

 
To better understand our population in terms of their mental health and wellbeing status, we administered the Depression Anxiety Stress Scales (DASS) \cite{lovibond1995structure}. Overall, the participant population was generally healthy in terms of their mental health scores, as measured by DASS. This questionnaire includes a set of self-report scales designed to measure negative emotional states of depression, anxiety, and stress. We utilized the short version of DASS, which includes 21 items, 7 per scale. Each item is rated on a Likert scale, ranging between 0 (never) to 3 (almost always). Total DASS has possible scores of 0-63, and depression, anxiety, and stress sub-scales have possible scores of 0-21. See Table \ref{tab:groupStats} for baseline values and their standard deviation among participants. Values under 4.5 for depression scale, under 3.5 for anxiety scale, and under 7 for stress scale are considered in the normal range.

 \begin{table}[h!]
   \centering
   \begin{tabular}{ l   p{1.5cm} p{1.5cm} p{1.5cm}}
     \multirow{2}{*}{\textbf{Group}} & \multicolumn{3}{c}{\textbf{DASS}} \\
      & Depression & Anxiety & Stress\\
     Emotion-Aware & 3.6$\pm$4.1 & 2.3$\pm$3.3 & 4.7$\pm$3.6 \\
     Control & 3.4$\pm$4.0 & 2.1$\pm$2.0 & 4.4$\pm$3.1 \\
   \end{tabular}
   \caption{Participants' mental health state as measured by DASS. Mean and standard deviation of baseline values for depression, anxiety, and stress sub-scales are included.}
     \label{tab:groupStats}
     \vspace{-.5cm}
 \end{table}

\section{Experiment: the Influence of Interacting with the Emotion-Aware Chatbot}\label{sec:experiment1}

Our research question is regarding the influence of interacting with an emotionally expressive bot compared to a neutral agent. Previous research has shown that interacting with a textual agent that shows minimal support of affect already helps to relieve strong negative affect. Also, when combined with a system that is designed to be frustrating, i.e., a game with unexpectedly long delays, participants prefer to continue to use such a system for longer if they are interacting with the emotional bot \cite{klein2002computer}. Subtle emotional expressiveness in agents has also been associated with higher trust and likability \cite{bickmore2003subtle}. Other researchers have looked into the role of personality (introversion/extraversion dimension) in interacting with virtual agents \cite{reeves1996media, nass2000does, buisine2010influence}. Building upon previous research, we would like to explore the following questions: Does interacting with the Emotion-Aware chatbot improve users' self-reported mood? Do extraverts benefit more from adding emotional expressiveness to bots compared to introverts?

To answer these questions, we designed a one-week, longitudinal experiment. We randomized participants into two groups: Emotion-Aware and Control. The Emotion-Aware group had access to the mobile app that administered experience sampling. The app would generate 5 probes at random times throughout the day, between 9AM and 9PM, approximately every 2.5 hours, and we made sure that the probes were at least 30 minutes apart. Each experience sampling prompt started with a phone notification from the app, saying ``Hi! Have a minute?''. The participants could then click on the notification, or start the app by clicking on the application icon on the home screen. After the app opened, the chatbot would randomly select from a set of initial prompts that asked the participant to report his/her emotional state. Then, it would provide the experience sampling visual grid (See Fig.~\ref{fig:UI}). After the participant responded to the prompt by dragging the indicator to express his/her emotional state, the chatbot would detect the selected quadrant, and randomly draw from a set of emotionally relevant phrases scripted for the respective quadrant. Note that the Control group had access to a similar interface, with the same methodology in triggering experience sampling probes. However, the responses to the experience sampling would always be selected from a pool of plain neutral texts without any expressive emotions. In summary, the difference between Emotion-Aware and Control users was in the responses that the participants received after reporting their mood. In the Control group, the app was only an instrument to capture data. Regardless of the user's selection, it would thank the user politely afterwards with a neutral tone; but in the Emotion-Aware group, the app would acknowledge the user's current status, respond appropriately, and resemble an empathetic companion.

\subsection{Measures}

To test our hypotheses regarding the interplay between personality and agent likability, we captured personality traits in the pre-study survey. We used well-validated measures of affect in the pre- and post-study surveys to capture affect and further validate the experience sampling data. We introduced satisfaction measures to study agent likability and user experience. Also, we analyzed the momentary mood sampled by the bot.

\subsubsection{Big Five Personality Traits}

The Big Five Personality Trait scale is a model based on common descriptors of personality that includes five factors: openness to experience, conscientiousness, extraversion, agreeableness, and Neuroticism \cite{digman1990personality}. The scale is composed of 44 items, where each item is rated on a Likert scale, ranging between 1 (strongly disagree) to 5 (strongly agree). Each personality factor is associated with 8-10 questions, thus possible scores are between 8-50.

\subsubsection{Positive and Negative Affect Schedule}

The Positive and Negative Affect Schedule (PANAS) consists of 20 words that describe different emotions \cite{watson1988development}. Half of the items indicate positive affect (PA) and half indicate negative affect (NA). Items are rated on a Likert scale, ranging from 1 (very slightly or not at all) to 5 (extremely). PA and NA are calculated separately and each range between 10-50. the PA/NA ratio is another commonly used measure derived from PANAS. PANAS has been used to capture affect in different time scale ranges. These include momentary, daily, over the past few days, weekly, for the past few weeks, yearly, and general affect. In our study, we have used PANAS to capture affect over the past week.

\subsubsection{User Preference}

We assessed satisfaction and efficacy of the system through different questions using a Likert scale, ranging from 1 (strongly disagree) to 7 (strongly agree). These questions asked about the agent's likability, intelligence, and the appropriateness of its ``tone". Questions were also asked about user preference for continuing to interact with the agent, and his/her improvement in awareness of daily emotions. They also asked if the notifications from the app where too frequent. Also, we included an open-ended question at the end of the week for general comments. 

\subsubsection{Experience Sampling}

Using the visual experience sampling grid, we capture valence ($v$) and arousal ($a$) on a continuous scale, $v, a\in [0.0,1.0]$. To match the discrete categories of the pool of scripted text explained in Section \ref{sec:scripted-text}, we discretize $v$ to have positive and negative valence:

\begin{equation}
\hat v= \left\{ \begin{array}{rcl}
Negative & \mbox{for}
& v<0.5 \\ Positive & \mbox{otherwise} & \\
\end{array}\right.
\end{equation}

We also discretize $a$ to have high and low arousal. 

\begin{equation}
\hat a = \left\{ \begin{array}{rcl}
Low & \mbox{for}
& a<0.5 \\ High & \mbox{otherwise} & \\
\end{array}\right.
\end{equation}

The 4 possible combinations of $\hat v$ and $\hat a$ are mapped to the 4 quadrants on the visual grid: Top Left (TL), Top Right (TR), Bottom Left (BL), and Bottom Right (BR).

\subsection{Results}

\subsubsection{User Perception of the Emotion-Aware Chatbot}

\begin{figure}[t!]
\centering
  	\includegraphics[width=0.8\columnwidth]{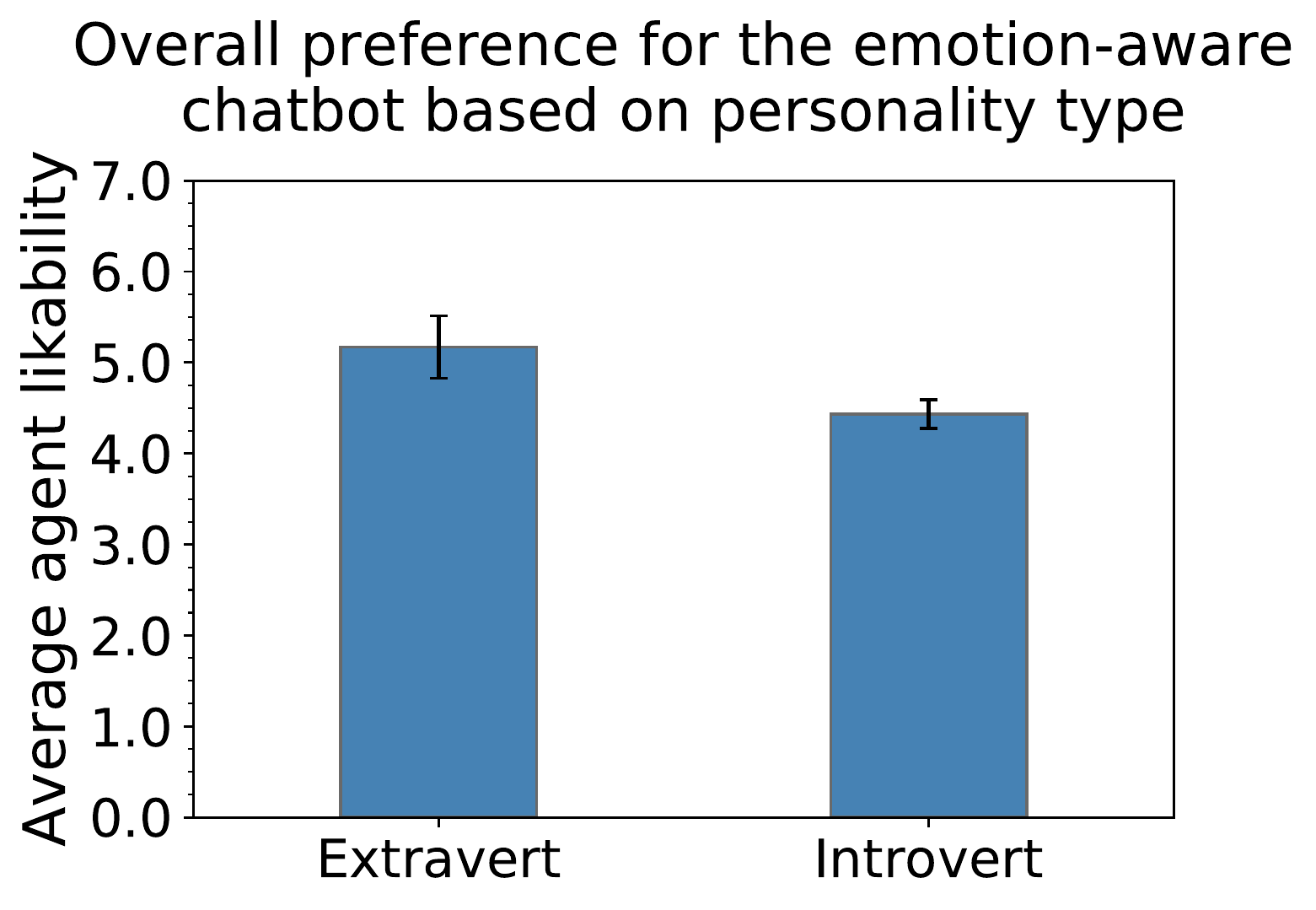}
  \caption{Overall agent preference based on personality type with one standard error bars. Extraverts preferred the Emotion-Aware chatbot significantly more.}
   \label{fig:extroversion}
   \vspace{-.4cm}
\end{figure}

Given that the Emotion-Aware chatbot is emotionally expressive, we questioned whether different personality types would prefer the agent more or less. Specifically, do extraverts prefer the Emotion-Aware chatbot more than intraverts? To answer this question, we discretized the Big Five extraversion scores into binary values: extravert (above median) vs. introvert (below median). Focusing only on the Emotion-Aware group, we compared the overall likability of the agent as averaged across all likability questions. An independent-samples t-test showed a significant difference in the overall likability scores for extraverts (M=5.17, SD=.91) and introverts (M=4.43, SD=.55); t(17)=2.08, p=.05 (Fig.~\ref{fig:extroversion})\footnote{The Pearson correlation coefficient and p-value between extraversion score and agent likability are the following: r=0.432, p=0.065.}.

\subsubsection{Influence of the Emotion-Aware Chatbot on Mood Reports}

\begin{figure}[t!]
\centering
  	\includegraphics[width=1.05\columnwidth]{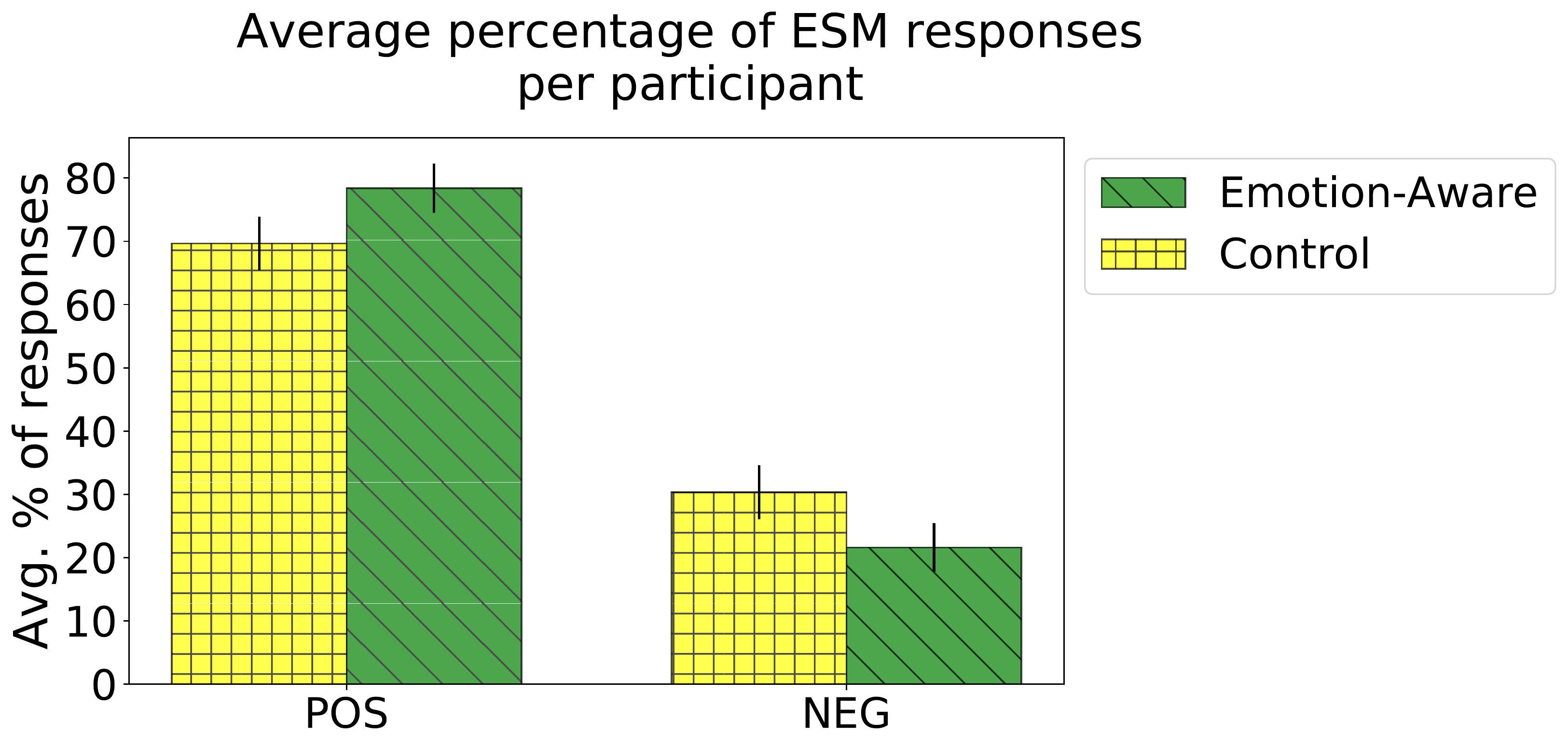}
  \caption{Average percentage of ESM responses per participant with one standard error bars. POS: positive valence quadrants on the 2x2 Russell circumplex model of emotion (top right and bottom right), NEG: negative valence quadrants (top left and bottom left). Participants in the Emotion-Aware condition reported higher percentage of positive emotions.}
   \label{fig:ESM-w1}
   \vspace{-.5cm}
\end{figure}

To answer this question, we compared the daily percentage of positive and negative ESM mood reports across groups. Granular daily self-reported emotion samples revealed significant differences across Emotion-Aware and Control groups. Fig.~\ref{fig:ESM-w1} shows the average percentage of the positive and negative ESM self-reports per participant. An independent-samples t-test was conducted to compare percentage of positive emotions reported daily between the Emotion-Aware and the Control conditions. There was a significant difference in the percentage of positive emotions for Emotion-Aware (M=80.55, SD=3.65) and Control (M=69.08, SD=4.16) conditions; t(37)=2.74, p = .009 \footnote {Since the percentage of negative emotions is 100 minus the percentage of positive emotions, similarly there was a significant difference in the percentage of negative emotions for Emotion-Aware (M=19.45, SD=3.65) and Control (M=30.92, SD=4.16) conditions; t(37)=-2.74, p = .009.}. The Emotion-Aware group reported a higher percentage of positive emotions (Top Right and Bottom Right affective quadrants in Russell's 2x2 model) and a lower percentage of negative emotions (Top Left and Bottom Left quadrants) compared to the Control group. Note that the weekly PANAS survey and the daily ESM are capturing instantaneous vs. weekly mood which are different by definition, but our analysis showed that the PA score derived from PANAS and the total number of positive self-reports over the course of the week were correlated (Pearson r=0.217, p=0.020). However, looking more closely at the influence of the Emotion-Aware chatbot on PA from weekly PANAS scores, a 2 (group) x 2 (pre-post PA) RM-ANOVA did not show a significant group x pre-post interaction \footnote{F=2.430, p=.098}.

\subsubsection{User Feedback}

Several participants reported interacting with the app as ``an interesting experience'' (pa070), ``pretty quick'' (pa081) and ``fun'' (pa045).

Some mentioned that the experience sampling made them more self-aware, or amplified their emotional state; pa050: ``notifications from the agent amplified how I was feeling.''; pa064:  ``It is a good exercise to periodically reflect on my emotions. I really like that aspect.''; pa063 mentioned surveys acted as a feedback loop, too: ``answering this survey forces me to define an emotional profile, to which I somehow become committed or identify with, which in turn influences my daily ratings.''.

Originally, we did not fully absorb the extensive role of the bot on self-awareness, but the overwhelming feedback from participants recognizing how it influenced their behavior highlighted that any behavior change application needs to support self-reflection. This result is in line with previous research findings, suggesting that self-reflection is an important part of behavior change and has the potential to improve wellbeing and mood \cite{isaacs2013echoes, taber2018personality}. However, it is worth mentioning that encouraging users to self-reflect should be done in moderation. There are downsides with interrupting users too frequently to self-report. First, the possible consequences should be considered. Some participants mentioned that an extremely high frequency of self-reflection could be harmful in certain circumstances; pa050: ``When I was stressed/worried at work and saw that I had to report on my feelings then those feelings felt more intense.''; pa088: ``I'm not sure if thinking about my feeling so many times in a day is a good thing. I realized that I've been picking happy only infrequently, which made me a little sad.''. Second, there can simply be high missing data rate. pa011: ``I frequently miss notifications.''. Therefore, if the application is solely relying on users' self-reported data, it will significantly hurt performance. Partial or full automation could help address these caveats.

Also, the open responses shed light on what could be improved. Some participants mentioned the chatbot's responses in the Emotion-Aware condition were exaggerated and unable to capture subtle or nuanced emotional states; pa073: ``The agent's responses seemed very \textit{narrow} responding with just a few generic phrases to my self-assessments[...]. Although the emotion quadrant consists of four squares, the actual coordinates within each square have a wide range of meanings [...]. However, the agent did not appear to respond particularly differently [to the intensity of the reported emotion].''; pa028:  ``The reactions to input could be better. They seem to come from the four basic zones (+- x/y), and the feedback from the bot doesn't indicate that the input I send is any more granular than that.''; pa067: ``The agent seems to respond as if your emotional state is either great or terrible[...]. It would be nice if it could adopt a more neutral tone in some circumstances. It's just kind of weird when it says something like \textit{Bummer} when I report that I'm feeling [almost] neutral.''; pa031: ``It would be nice if the agent had better design and some kind of persona. The way it is now seems simplistic (though, still useful in the sense of reminding)''.

We need to emphasize that there was overwhelming feedback from participants highlighting that they wanted to be able to enter more nuanced emotional state self-report data, but that the bot's responses were coarse and rough and did not account for the subtleties in their reports. In other words, users wanted to select very particular feelings during self-report, and they wanted an agent that reflected that level of precision. This lack of granularity bothered our users, even though they still liked the reminding facility.

Some of the responses mentioned the difficulty users had in expressing precise emotional reports. This could be due to the UI; pa078: ``It's difficult to be precise in positioning the dot on the axes.''. It could also be due to difficulty identifying emotions and mapping them to the quadrants; pa061: ``Sometimes I found it hard to describe my feelings'', pa052: ``[the] subtle changes in my emotions are not being captured by my current way of recording it.'' Providing better user interface support for users to reflect upon and enter their emotions remains an important issue for future work.

\section{Discussion}\label{sec:discussion}

\subsection{Empathetic Experience Sampling and Mood}

Our results showed that providing an emotionally appropriate response when conducting experience sampling, similar to what happens in a successful human-human interaction, resulted in a higher percentage of positive responses being recorded. However, interaction with the agent did not significantly influence positive and negative affect, as captured by the weekly PANAS surveys. We have three possible interpretations.
First, the influence of the agent may be subtle and, since it only appeared in granular experience sampling about five times a day, was possibly not enough to show its influence over one week. 
Second, the emotional expressiveness of the bot may have resulted in self-report bias. One participant said: "Sometimes, the responses when the mood is marked as negative seem somewhat validating or disheartening, subconsciously making me reluctant to mark my mood as such." This suggests that the affirmative response from the agent might have affected the ratio of missing self-reports asymmetrically for negative vs. positive samples. 
Third, our population was overall quite healthy and happy--improved positive affect in a clinical sense would probably be unlikely. Further studies are needed to get a deeper understanding about empathetic experience sampling to tease these issues apart.

\subsection{Personality and Preference for an Affective Agent}

Our findings revealed that there is value in adding emotional understanding and expression to conversational agents. The emotionally expressive bot was generally liked (on average more than 4 from on a 7-scale Likert scale). However, extraversion was an important personality factor influencing the likability of the agent: extraverts' average likability measures were significantly higher than introverts'. This suggests that certain personality types may benefit more from adding emotional intelligence or expressiveness to conversational agents.

\subsection{Design Guidelines}

A synthesis of users' feedback shed light on guidelines that could prove useful for designing affective conversational bots. 

\textit{Emotional intelligence is sometimes a neutral response.}
Feedback from participants revealed that providing emotionally expressive responses to subtle emotions decreased the perception of emotional intelligence of the bot. For example, expressing sympathy in response to minor expressions of sadness was received as unnecessary exaggeration. Instead, a neutral or nuanced response was preferred. We learned that low intensity emotions should be responded to with more subtle and neutral interactions.

\textit{Behavior change applications benefit from supporting self-reflection.}
The overwhelming feedback from our participants shed light on the influence of self-reflection on behavior change. We suggest that any behavior change application should consider supporting self-reflection to improve the efficacy of the system. We need to highlight that supporting self-reflection does not necessarily require sole reliance on the user to provide data frequently. It rather means intelligent support systems could provide opportunities for the user to self-reflect at the right pace and frequency, while still being able to function without needing high rates of user data.

\subsection{Limitations}

We manually scripted all the textual interactions. Though we created multiple phrases with similar, but slightly different messages, their occurrence soon became ``expected" over the course of the study. In the future, we would like to use machine learning to automate the intervention text generation and make it emotionally expressive by adding emojis or sentiment that works for an individual according to personal preference and context \cite{felbo2017using, hu2017toward, ghandeharioun2019approximating}. 
This work lies on the boundary between a data-gathering and a behavior change tool. It is worth mentioning that the emotional expressiveness of the Emotion-Aware bot may have confounded the experience being sampled. Further studies are required to fully investigate validity of ESM responses, with and without these modifications.
Further replication of this work using larger populations is encouraged.

\section{Conclusions and Future work}\label{sec:conclusions}

In this paper, we quantitatively and qualitatively evaluated an emotion-aware chatbot for conducting experience sampling, over the course of a week, with N=39 participants. Our results show that an emotionally expressive agent is likable, particularly to extraverts. Furthermore, an emotionally sentient agent such as we deployed has the potential to improve positive affect and reduce negative affect.
We identified several design guidelines for future work. Specifically, we found that an emotionally appropriate response is sometimes neutral and that support of self-reflection is a crucial part of behavior-change applications.
In the future, we would like to extend our system to detect a user's mood from passive smartphone sensor data and use automatically predicted emotional states to drive emotional dialog and relevant micro-interventions just-in-time. 

\bibliographystyle{IEEEtran}
\bibliography{EMMA}

\end{document}